\documentclass[pre,onecolumn,superscriptaddress]{revtex4-1}
\usepackage{latexsym}
\usepackage{amsmath}
\usepackage{amsthm}
\usepackage{times}
\usepackage{amssymb}
\usepackage{fancyheadings}
\usepackage[T1]{fontenc}
\usepackage[utf8]{inputenc}
\usepackage{float}
\usepackage[english]{babel}
\usepackage{fancyhdr}
\usepackage{url}
\usepackage{hyperref} 
\usepackage{color}
\usepackage{graphicx} 
\graphicspath{{Figures/}}
\usepackage{bm}
\usepackage[capitalize]{cleveref} 
\usepackage[section]{placeins} 

\date{Jul. 20, 2021}

\theoremstyle{theorem}

\theoremstyle{definition}

\theoremstyle{lemma}

\Crefname{equation}{Eq.}{Eqs.}
\Crefname{figure}{Fig.}{Figs.}
\Crefname{tabular}{Tab.}{Tabs.}

\begin{abstract}
In this work, we introduce a new methodology for inferring the interaction structure of discrete valued time series which are Poisson distributed. While most related methods are premised on continuous state stochastic processes,  in fact, discrete and counting  event oriented stochastic process are natural and common, so called time-point processes (TPP). 
An important application that we  focus on here is gene expression.  Nonparameteric methods such as the popular k-nearest neighbors (KNN) are slow converging for discrete processes, and thus  data hungry. Now, with the new multi-variate Poisson estimator developed here as the core computational engine, the causation entropy (CSE) principle, together with the associated greedy search algorithm optimal CSE (oCSE) allows us to efficiently infer the true network structure for this class of stochastic processes that were previously not practical.  We illustrate the power of our method, first in benchmarking with  synthetic datum, and then by inferring the genetic factors network from a breast cancer micro-RNA (miRNA) sequence count data set. We show the Poisson oCSE gives the best performance among the tested methods anfmatlabd discovers previously known interactions on the breast cancer data set. 
\end{abstract}

\begin{document}

\title{Interaction Networks from Discrete Event Data by Poisson Multivariate Mutual Information Estimation and Information Flow with Applications from Gene Expression Data}
\author{Jeremie Fish}
\affiliation{Department of Electrical and Computer Engineering Clarkson University}
\affiliation{Clarkson Center for Complex Systems Science}
\email{fishja@clarkson.edu}
\author{Jie Sun}
\affiliation{Clarkson Center for Complex Systems Science}
\email{sunj@clarkson.edu}
\author{Erik Bollt}
\affiliation{Department of Electrical and Computer Engineering Clarkson University}
\affiliation{Clarkson Center for Complex Systems Science}
\email{bolltem@clarkson.edu}

\maketitle
\section{Introduction}
Understanding the behavior of a complex system requires knowledge of its underlying structure. However prior knowledge of the network of interactions is often unavailable, necessitating estimation from data.
Perhaps no complex system is more important to our health and well being  than   
Data-driven analysis of 
gene expression is a complex system that is especially important to our health and well being.
However, these data are generally  time-point process (TPP), and discretely distributed,  rather than continuous valued as most mutual information inference methods  presume. Specifically, we assume a jointly distributed Poisson process.  While TPP are relatively common, as far as we know, no efficient joint entropy estimator exists. To this end, the main goal of this paper is to fill that gap. 
\paragraph*{}

Granger causality \cite{granger1969} has been used for
network inference when interpreted as a causation inference concept, for linear stochastic processes, as well as transfer entropy (TE) \cite{schreiber2000} based on information theorety for nonlinear processes. Howeverm when applied to a system with more than two factors, neither of these concepts can distinguish direct versus indirect effects or co-founders, and therefore they will tend to yield false positive connections.  To this end, we developed causation entropy (CSE) as a generalization of transfer entropy \cite{sun2014,sun2015}, that explicitly defines the information flow between two factors, conditioned on  tertiary intermediary factors. This, together with a greedy search algorithm to construct the network of interactions of the complex stochastic process, provably reveals network structure of certain stochastic processes, \cite{sun2015}.  In past studies, TE as well as CSE were computed nonparametrically, by the  Kraskov-St\"ogbauer-Grassberger (KSG) \cite{kraskov2004}  mutual information estimator  which is a K-nearest neighbors (KNN) method.  However, if specific knowledge of the joint distribution of the process allows considerable computation efficiencies, such as our previous work where jointly Gaussian variables \cite{sun2015} or jointly Laplace distributed  variables  in \cite{ambegedara2016} were relevant.

\paragraph*{}
Here, we focus on  gene expression networks, which are an application of considerable scientific importance due to their foundational relevance as a building block tool to understanding details of life science. 
It is well understood that many diseases  associate with  variations of the expression of a single gene \cite{rogers2008,sebastiani2005,deboulle1993}, e.g., famously such as  sickle cell disease and cystic fibrosis. However it remains a difficult problem with considerable health implications to explain and to infer complex interactions and associations when many genes may be involved in common and even deadly disease.  Such diseases are called polygenetic, and these include the breast cancer example that we study here.  According to the Centers for Disease Control (CDC), in the USA, breast cancer is considered to be the second most common form of cancer amongst woman, \cite{CDC2019}, that in 2019 was forecast to 268,600 cases and 42,260 deaths in 2019.  
We advance here a new methodology to probe variations in expression of a group (network) of genes that may lead to disease. Understanding the gene interaction network structure may be crucial to the development of future treatments.  Network inference itself has many applications beyond cancer research, including fMRI network inference \cite{smith2012, bassett2011,stoltz2017, fish2021}, drug-target interaction networks \cite{yamanishi2008}, and earthquake network inference \cite{zhang2016} and economy issues \cite{iori2008} to name a few.

\paragraph*{}
With this motivation, the main technical premise of this paper is to develop a computationally efficient approach to estimate joint entropy and related information theoretic measures for multivariate Poisson processes.  Data derived from these are discrete-valued data, and typically  consist of   a significant fraction of zeros punctuated with nonzero values describing event counts in a given epoch. From the multi-variate Poisson model, we derive an analytical series representation of the joint entropy and the mutual information. Then, a practical finite partial-sum estimator allows estimation of  mutual information, toward transfer entropy and causation entropy. 

\paragraph*{}
This paper is structured as follows: first, we provide a brief introduction to  mathematical background including a multivariate Poisson model and also relevant information theoretic quantities which are necessary to define information flow. Then, we  derived our multivariate Poisson joint entropy estimator, which we relate to  network inference. Finally, in the Results section we demonstrate our method and performance for benchmark synthetic data  and then we study the  breast cancer gene expression data sets.

\begin{figure}[htbp] 
\includegraphics[trim = 0 250 0 0,width=0.9\textwidth]{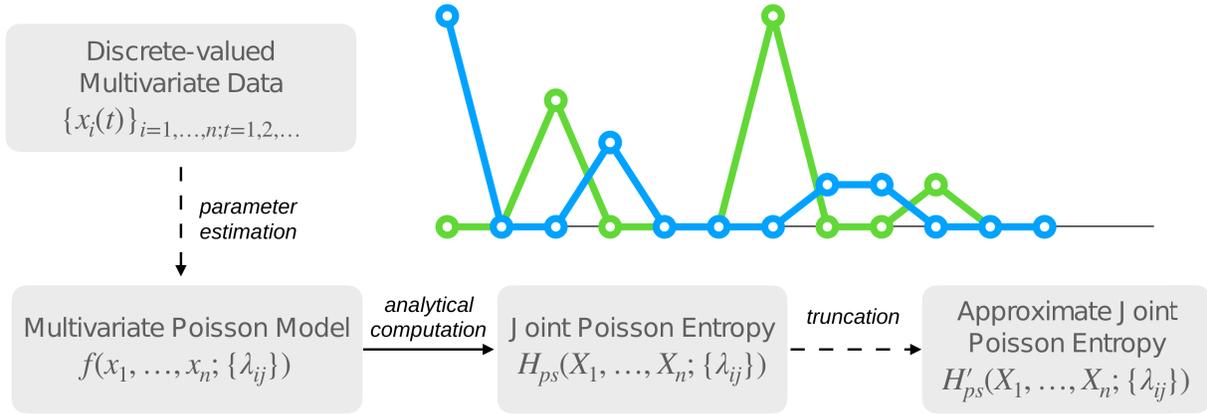} 
\caption{Work flow of our computationally efficient approach to estimate the joint entropy of multi-variate Poisson distributed variables. From data, we proceed to distribution parameter estimation to approximate joint entropy.  \label{fig:figure1}}
\end{figure} 
\section{Background}
\subsection{Multivariate Poisson Model}
First let us recall the single variate Poisson Model,  \cite{reiss2012,bollt2013}:
\begin{equation}
p(k) = \frac{\lambda^k}{k!}e^{-\lambda}. \label{eq:svPoisson}
\end{equation}
The Poisson model has a multivariate generalization as follows, \cite{karlis2007}:
\begin{align}
& P(X_1 = x_1, ... ,X_n = x_n)  = \label{eq:multiPoiss_a}  \\ \nonumber
& e^{-\sum \limits_{i=1}^n \sum \limits_{j\geq i} \lambda_{ij}} \sum \limits_\mathcal{C} \frac{\prod \limits_{i=1}^n \lambda_{ii}^{(x_i - \sum \limits_{j}a_{ij})} \prod \limits_{i = 1}^n \prod \limits_{j > i} \lambda_{ij}^{a_{ij}} }{\prod \limits_{i = 1}^n (x_i - \sum \limits_{j}a_{ij})! \prod \limits_{i = 1}^n \prod \limits_{j > i} a_{ij}!}, 
\end{align}
where the set
\begin{align}
&\mathcal{C} \label{eq:multiPoiss_b}= \\ \nonumber
&\{A = [a_{ij}]_{n \times n}| a_{ij} \in \mathbb{N}_0, a_{ii} = 0, a_{ij} = a_{ji} \geq 0, \\ \nonumber
& \sum \limits_j a_{ij} \leq x_i \},
\end{align}
and $\mathbb{N}_0 = \mathbb{N} \cup \{0\}$. This model is based on assuming that the $x_i$ are linearly transformed from a set of independently drawn Poisson variables. We begin with
\begin{align}
&X \in \mathbb{N}_0^{n \times t} = (x_1,x_2,...x_n)^T  = BY, \label{eq:multiPoiss_c} \\ \nonumber
&Y \in \mathbb{N}_0^{m \times t} = (y_{11},y_{22}, ..., y_{nn}, y_{12}, y_{13},...,y_{(n-1)n})^T.
\end{align}
 Here each $y_{ij}$ is independent Poisson, that is: $y_{ij} \in \mathbb{N}_0^t \sim \mbox{Poisson}(\lambda_{ij}), (\mbox{for} \ i = 1,...,n, j \geq i)$, so $m = n+ \frac{n(n-1)}{2}$, $B \in \mathbb{N}_0^{n \times m}$. Note that in this case $\lambda_{ij} = \lambda_{ji}$. The rows of $X$ thus represent Poisson random variables which have $t$ observations. Although the number of parameters needed to specify this model grows quickly, there are some nice properties. For instance, this model allows a simple estimate of  each $\lambda_{ij}$, since the sum of independent Poisson variables yields the following covariance matrix structure:
\begin{equation}
\mbox{Cov}(X) =
\begin{bmatrix}
\lambda_{11} + \sum \limits_{ j\neq 1}^n \lambda_{1j} & \lambda_{12} & \cdots & \lambda_{1n} \\
\lambda_{12} & \lambda_{22} + \sum \limits_{j\neq 2}^n & \cdots & \lambda_{2n} \\
\vdots & \vdots & \cdots & \vdots \\
\lambda_{n1} & \lambda_{n2} & \cdots  & \lambda_{nn}  + \sum \limits_{j = 1}^{n-1} \lambda_{nj}
\end{bmatrix} \label{eq:CovarianceMat}
\end{equation}
with $\lambda_{ij} = \lambda_{ji}$. The $(i,j)$ entries of the covariance matrix represent $\mbox{cov}(x_i,x_j)$,  the covariance between the two random variables, $x_i$ and $x_j)$. Proof of Eq. \ref{eq:CovarianceMat} may be found in the appendix. 

This model is a multivariate extension of the Poisson model that does not assume the random variables are necessarily independent. However, there are some limitations to this model. First, the rapid growth in the number of states and parameters with respect to the number of variables, making calculation of the joint distribution computationally unwieldy and expensive. Another limitation is that  model cannot handle negative covariance \cite{karlis2007}. These difficulties are particularly complicating in the forth coming entropy computations, and so they must  be handled in later sections.
\subsection{Transfer Entropy and Causation Entropy}
We  briefly review certain Shannon entropies, building toward the concepts of transfer entropy and causation entropy. These are the fundamental concepts of information flow we use to consider network inference. The Shannon {\it entropy} of a (discrete) random variable $X$ is given by \cite{shannon1948,cover2012}:
\begin{equation}
H(X) = - \sum \limits_{x \in X} P(x)\mbox{log}(P(x)) \label{eq:shannon_entropy},
\end{equation}
where $P(x)$ is the probability that $X = x$, and $0 \mbox{log}(0) = 0$ is the usual interpretation in this context. For the remainder of this paper we choose the natural log and thus all entropies will be measured in nats. Entropy can be thought of as a measure of how uncertain we are about a particular outcome. As an example we can imagine two scenarios, in one case we have a random variable $X_1 = (x^{(1)}_1, x^{(2)}_1,...,x^{(n)}_1)$ with $x^{(t)}_1 = 0 (\forall t)$, that is $P(X_1=0) =1$,  in the other case the random variable $X_2 = (x^{(1)}_2, x^{(2)}_2,...,x^{(n)}_2)$ with $P(X_2= 0)=0.5, \ P(X_2= 1) = 0.5$. Here $H(X_1) = 0$ nats, while  $H(X_2) = \mbox{ln}(0.5)$ nats which happens to be the maximum for this case \cite{cover2012}. It is easy to see that  Shannon entropy reaches it's greatest value when we are the most uncertain about the outcome, and its minimal value ($0$) when we are completely certain about the outcome. We can now examine the case of two random variables $X$ and $Y$. The {\it joint entropy} of a discrete random variable is defined \cite{cover2012}:
\begin{equation}
H(X,Y) = - \sum \limits_{x \in X} \sum \limits_{y \in Y} P(x,y) \mbox{ln}(P(x,y)). \label{eq:joint_entropy}
\end{equation}
When the two random variables $X$ and $Y$ are independent $H(X,Y) = H(X) + H(Y)$ which is the maximum joint entropy. Thus $H(X,Y) \leq H(X) + H(Y)$. There are comparable definitions of differential  entropies for  continuous random variables in terms of integration. The {\it conditional entropy} is defined:
\begin{equation}
H(X|Y) = - \sum \limits_{x \in X} \sum \limits_{y \in Y} P(x,y)\mbox{ln}P(x|y).
\end{equation}
The conditional entropy gives us a way to describe the relationship between variables, which is the key to network inference. If knowledge of the variable Y gives us complete knowledge of the variable X then the conditional entropy will be $H(X|Y)= 0$ nats. 
Another important Shannon entropy is the {\it mutual information} which is defined as \cite{cover2012}:
\begin{align}
I(X,Y) =  \sum \limits_{x \in X} \sum \limits_{y \in Y} P(x,y) \mbox{ln}\Big(\frac{P(x,y)}{P(x)P(y)}\Big) =& \nonumber \\
H(X)-H(X|Y) =& \nonumber \\
H(X)+H(Y)-H(X,Y). \label{eq:mutual_information}
\end{align}
 Finally, the {\it Kullback-Leibler (KL) divergence} ($D_{KL}$) \cite{cover2012} is stated:
\begin{equation}
D_{KL}(P||Q) = - \sum \limits_{x\in X} P(x) \mbox{ln}\Big(\frac{Q(x)}{P(x)}\Big). \label{eq:KL}
\end{equation}
The KL divergence describes a distance-like quantity between two probability distributions, though it is not a metric as for one, it is not symmetric (that is in general $D_{KL}(P||Q) \neq D_{KL}(Q||P)$), and  also it does not satisfy the triangle inequality. Mutual information Eq. \ref{eq:mutual_information} can be written in terms of KL divergence as \cite{cover2012}:
\begin{equation}
I(X,Y) = D_{KL}(P(x,y)||P(x)P(y)),
\end{equation}
describing a deviation from independence of a joint random variable $(x,y)$.
For a stationary stochastic process, $\{ X^t \}$, the {\it entropy rate} is defined as \cite{cover06, bollt13}: 
\begin{equation}
H(\chi) = \lim \limits_{t \rightarrow \infty} H(X^t | X^{t-1}, X^{t-2}, ... X^1).
\end{equation}
If the process is Markov (memoryless) then \cite{cover2012}:
\begin{equation}
H(\chi) = \lim \limits_{t \rightarrow \infty} H(X^t|X^{t-1}).
\end{equation}
The {\it transfer entropy} from $X_2$ to $X_1$ is defined \cite{schreiber2000, sun2014}:
\begin{equation}
T_{X_2 \rightarrow X_1} = H(X^{t+1}_1|X^t_1) - H(X^{t+1}_1|X^t_1,X^t_2).
\end{equation}
{\it Causation entropy} is  a generalization of the transfer entropy, where \cite{sun2014,sun2015}:
\begin{equation}
C_{\mathcal{Q} \rightarrow \mathcal{P}|\mathcal{S}} = H(\mathcal{P}^{t+1}|\mathcal{S}^t) - H(\mathcal{P}^{t+1}|\mathcal{S}^t,  \mathcal{Q}^t). \label{eq:causationentropy}
\end{equation}
\begin{figure}[htbp] 
\includegraphics[width=0.65\textwidth]{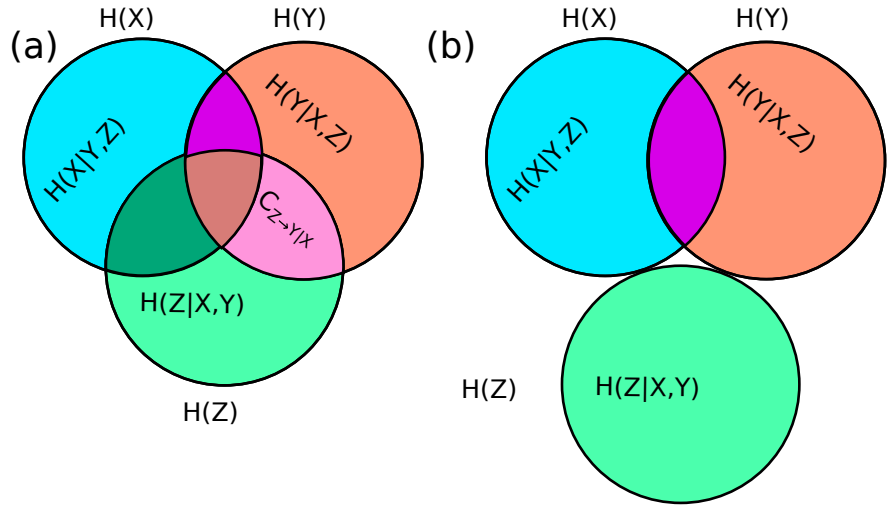}
\caption{(a) The causation entropy between two processes $Z$ and $Y$ is shown. In this case since we are only conditioning on a process $X$, $C_{Z \rightarrow Y|X} = T_{Z \rightarrow Y}$. Of course $X$ may be replaced with a set of variables. (b) Here we show a special case where $Z$ is independent of both $X$ and $Y$ ($Z$ in this case may represent the history of $X$. In this case it becomes clear that $H(Z|X,Y) = H(Z)$, $H(X|Y,Z) = H(X|Y)$ and $H(Y|X,Z) = H(Y|X)$. As explained in the text, this special case helps us to discern what are the proper variables to use in the Poisson case. \label{fig:informationtheory}}
\end{figure} 
$C_{\mathcal{Q} \rightarrow \mathcal{P}|\mathcal{S}}$ is designed to describe the remaining information flow from processes $\mathcal{Q}$ to processes $\mathcal{P}$ that may not accounted for (conditioned on) processes $\mathcal{S}$.  An example of causation entropy is shown in Fig. \ref{fig:informationtheory} (a). In theory if a process $Z$ has no influence over another process $Y$, the causation entropy after conditioning out the remaining processes would be identically $0$, allowing us to reject a connection from $Z$ to $Y$. In practice however, when estimating these quantities by statistics from finite samples of noisy data, these will not compute to be identically $0$, making it necessary to have a threshold, which is the purpose of using a shuffle test as discussed in \cite{sun2015}. 

Network inference can be developed based on 
Eq.~(\ref{eq:causationentropy}).  However, considering the power-set of all possible subsets $\mathcal{P}, \mathcal{Q}, \mathcal{S}$ is clearly NP-hard and so not practical.
This led to the development of a greedy search algorithm, we referred to as optimal causation entropy (oCSE) \cite{sun2015, ambegedara2016} to a minimal network that explains the data, in terms of minimal causation entropy.
This proceeds in two stages, aggregative discovery of statistically significant links, those that are maximally informative influencers in terms of the conditionally already significant links, with possible removal of statistically irrelevant links developed while growing the global network, and significance decided by a null hypothesis in terms of multiple random shuffles of the data. We were able to prove under mild hypothesis of the stochastic process that this procedure will discover the true network, also assuming  a good statistical  estimation of the entropies.  It is precisely this problem of good data-driven statistical estimation of entropies specialized to the scenario of a multivariate Poisson process which is what we handle in this paper in the case of multivariate Poisson.

\section{Entropy Estimation from Multivariate Poisson Data}
\subsection{Estimating Joint Entropy of Poisson Systems}

Here we develop an estimator of entropies for the multivariate Poisson distribution, Eq.~(\ref{eq:multiPoiss_a}).  To this end, we truncate partial sums from series representations. 
\subsection{Poisson Entropy}
We begin the Poisson Entropy:
\begin{align}
\label{eq:poissonEntropy}
H_{Poisson}(K) = - \sum \limits_{k = 0}^\infty \frac{\lambda^k}{k!}e^{-\lambda} \mbox{ln} \Big(\frac{\lambda^k}{k!}e^{-\lambda} \Big) = \\
\nonumber  - \sum \limits_{k = 0}^\infty \frac{\lambda^k}{k!}e^{-\lambda} [- \lambda +k \mbox{ln}(\lambda) - \mbox{ln}(k!)] = 
\\
\nonumber \lambda - \lambda \mbox{ln}(\lambda) + \sum \limits_{k = 0}^\infty \frac{\lambda^k}{k!}e^{-\lambda} \mbox{ln}(k!). 
\end{align}
This expression for the entropy of a  Poisson random variable is in terms of an infinite series, which is well approximated by a finite truncation partial sum.  
\subsection{Bivariate Poisson Entropy}
The Bivariate Poisson case is instructive to the n-variate Poisson case. Consider: 
\begin{widetext}
\begin{equation} 
\label{eq:bivariatePoisson}
P(x_1,x_2) = e^{- \lambda_{11} - \lambda_{22} - \lambda_{12}} \frac{{\lambda_{11}}^{x_1}}{x_1!} \frac{{\lambda_{22}}^{x_2}}{x_2!} 
\bigg(\sum \limits_{a_{12} = 0}^{min(x_1,x_2)} \frac{x_1!}{(x_1 - a_{12})!} \frac{x_2!}{(x_2-a_{12})!a_{12}!}\Big(\frac{\lambda_{12}}{\lambda_{11} \lambda_{22}}\Big)^{a_{12}}\bigg).
\end{equation}
\end{widetext}
Let,
\begin{equation}
d_{12} = \frac{\lambda_{12}}{\lambda_{11}\lambda_{22}},
\end{equation}
and,
\begin{equation}
D(x_1,x_2) = \sum \limits_{a_{12} = 0}^{min(x_1,x_2)} \frac{x_1!}{(x_1 - a_{12})!} \frac{x_2!}{(x_2-a_{12})!} \frac{d_{12}^{a_{12}}}{a_{12}!}.
\end{equation}
Then Eq. \ref{eq:bivariatePoisson} will become:
\begin{equation}
P(x_1,x_2) = e^{- \lambda_{11} - \lambda_{22} - \lambda_{12}} \frac{{\lambda_{11}}^{x_1}}{x_1!} \frac{{\lambda_{22}}^{x_2}}{x_2!} D(x_1,x_2). 
\end{equation}
Now to get the joint entropy of the Bivariate Poisson we have:
\begin{widetext}
\begin{flalign}
H(X_1,X_2) &= -\sum \limits_{x_1 = 0}^\infty \sum \limits_{x_2 = 0}^\infty P(x_1,x_2) \mbox{ln}(P(x_1,x_2)) = -\sum \limits_{x_1 = 0}^\infty \sum \limits_{x_2 = 0}^\infty \nonumber \\
& e^{- \lambda_{11} - \lambda_{22} - \lambda_{12}} \frac{{\lambda_{11}}^{x_1}}{x_1!} \frac{{\lambda_{22}}^{x_2}}{x_2!} D(x_1,x_2) [-\lambda_{11} - \lambda_{22} -\lambda_{12} + x_1 \mbox{ln}(\lambda_{11}) + x_2 \mbox{ln}(\lambda_{22}) - \mbox{ln}(x_1!) - \mbox{ln}(x_2!) + \mbox{ln}(D(x_1,x_2))]. \label{eq:bivarEntropy}
\end{flalign}
\end{widetext}
A scenario of interest arises when $\lambda_{11},\lambda_{22},$ and $\lambda_{12}$ are all small and $\lambda_{12} << \lambda_{11}\lambda_{22}$. . In this case we have \begin{equation}
D(x_1,x_2) \approx \sum \limits_{a_{12} =0}^{min(x_1,x_2)} \frac{d_{12}^{a_{12}}}{a_{12}!},
\end{equation}
since the $d_{12}$ term dominates. Small $\lambda_{11}$ and $\lambda_{22}$  ensures that the large $x_1$ and $x_2$ terms to become insignificant in Eq. \ref{eq:bivarEntropy}. Thus, $D(x_1,x_2) \approx 1+\frac{d_{12}^2}{2!} +... \approx 1$. 
Grouping  terms and remembering (the middle part of) Eq. \ref{eq:poissonEntropy}, and estimating $D(x_1,x_2) =1$ , a finite partial sum of Eq. \ref{eq:bivarEntropy} can be written: 
\begin{equation}
H(X_1,X_2) = e^{-\lambda_{12}}[H(X_1) + H(X_2) + \lambda_{12}]. \label{eq:Estimate1}
\end{equation}
Remembering the assumption $\lambda_{12} << 1$, the expression Eq. \ref{eq:Estimate1} reduces further:
\begin{equation}
H(X_1,X_2) = [H(X_1) + H(X_2) + \lambda_{12}]. \label{eq:Estimate2}
\end{equation}
As Fig. \ref{fig:figure1} shows, this approximation works well when $d_{12} <<1 \implies \lambda_{12} << \lambda_{11} \lambda_{22}$, and in this regime the error will be small. Similar analysis can be carried out for the larger multivariate cases which allows us to arrive at a general formula for our approximation given by:
\begin{equation}
H(X_1,X_2,...,X_n) \approx [H(X_1) + ... H(X_n) + \sum \limits_{j>i} \lambda_{ij}], \label{eq:PoissonJointApprox}
\end{equation}
where we are assuming that $\lambda_{ij}$ are small for all $(i,j)$ pairs. Fortunately as we can see in Eq. \ref{eq:PoissonJointApprox}, all of the quantities on the right hand side are computationally efficient to compute. This in fact greatly reduces the computational time necessary for estimation of the joint entropy. This formulation  requires asymptotic assumptions that may not be valid in general in nature. However we find empirically in simulations that by scaling the rates $\lambda_{ij}$ to be in $[0, 1]$ the estimate performs well, as described by Fig. \ref{fig:figure1} and, verified in the network simulations, regardless of what the true underlying rates this scaling produces similar results.
\paragraph*{}
As a note of caution, consider that when calculating the mutual information in the Poisson model, care must be taken due to how the marginals of a joint Poisson process are drawn. For example from Eq. \ref{eq:mutual_information} it may be tempting to assert:
\begin{equation}
I_{Poisson}(X_1,X_2) = H(X_1)+H(X_2) - H(X_1,X_2), \label{eq:MIpoissonWrong}
\end{equation}
with $X_1 \sim \mbox{Poisson}(\lambda_{11})$ and $X_2 \sim \mbox{Poisson}(\lambda_{22})$. However this is not exactly correct, though the error here is subtle. In fact we must make a small change to Eq. \ref{eq:MIpoissonWrong} to be:
\begin{equation}
I_{Poisson}(X_1,X_2) = H(\hat{X_1})+H(\hat{X_2}) - H(X_1,X_2), \label{eq:MIpoissonCorrect}
\end{equation}
here $X_1 \sim \mbox{Poisson}(\lambda_{11})$ and $X_2 \sim \mbox{Poisson}(\lambda_{22})$, but $\hat{X_1} \sim \mbox{Poisson}(\lambda_{11}+\lambda_{12})$ and $\hat{X_2} \sim \mbox{Poisson}(\lambda_{22}+\lambda_{12})$. This subtle difference is important, because without recognizing this fact, the calculated mutual information becomes negative, which violates our well established condition that mutual information be positive. The need for $\hat{X_1}$ and $\hat{X_2}$ is apparent from Eq. \ref{eq:CovarianceMat}, when two Poisson random variables are summed together their {\it marginals} then are drawn from the sum of the underlying rate (i.e. $\lambda_{ii}$) and the coupling rate (i.e. $\lambda_{ij}$). This also transfers to computing the conditional mutual information. To better illuminate this calculation it is helpful to refer to Fig. \ref{fig:informationtheory} (b). 
\begin{equation}
I(X,Y|Z) = H(X,Z) + H(Y,Z) - H(X,Y,Z) - H(Z). \label{eq:3VariableCMI}
\end{equation}
In the special case presented in Fig. \ref{fig:informationtheory} (b) Eq. \ref{eq:3VariableCMI} becomes
\begin{equation}
I(X,Y|Z) = I(X,Y) = H(X) + H(Y) - H(X,Y), \label{eq:CMIspecialCase}
\end{equation}
therefore,
\begin{equation}
H(X) + H(Y) -H(X,Y) = H(X,Z) + H(Y,Z) - H(X,Y,Z) - H(Z). \label{eq:CMIspecialCase2}
\end{equation}
In this special case we can note the following:
\begin{equation}
H(Y,Z) = H(Y) + H(Z). \label{eq:CMIspecialCaseY}
\end{equation}
Applying Eq. \ref{eq:CMIspecialCaseY} to Eq. \ref{eq:CMIspecialCase2} we find that:
\begin{equation}
H(X) - H(X,Y) = H(X,Z) - H(X,Y,Z). \label{eq:CMIspecialCase_SolvedX}
\end{equation}
We know from Eq. \ref{eq:MIpoissonCorrect} that in the Poisson case this becomes:
\begin{equation}
H(\hat{X}) - H(X,Y) = H(X,Z) - H(X,Y,Z). \label{eq:CMIspecialCase_SolvedX_end}
\end{equation}
Applying the following facts to Eq. \ref{eq:CMIspecialCase_SolvedX_end}
\begin{equation}
\begin{cases}
H(X,Y,Z) = H(X,Y) + H(Z), \\
\mbox{and} \ H(X,Z) = H(X) + H(Z), \label{eq:CMIspecialCases}
\end{cases}
\end{equation}
we find that:
\begin{equation}
H(\hat{X}) = H(X).
\end{equation}
Similar analysis also shows that:
\begin{equation}
H(\hat{Y}) = H(Y),
\end{equation}
this implies that we must use the Poisson marginals in the computation of the conditional mutual information. That is in the Poisson case we must have: 
\begin{equation}
I(X,Y|Z) = H(\hat{X},Z) + H(\hat{Y},Z) - H(X,Y,Z) - H(Z). \label{eq:3VariableCMI}
\end{equation}
Note the use of $\hat{X}$ and $\hat{Y}$ in this case. This distinction in the Poisson case is important because we note that without using the proper marginals the computation results in {\it negative} conditional mutual information which is clearly not correct since conditional mutual information must be positive \cite{cover2012}.
\paragraph*{}
Importantly the new definition given in Eq. \ref{eq:PoissonJointApprox} becomes more computationally efficient than computing the Poisson joint entropy directly from the joint probability. This requires calculation of only separate {\it single} variate entropies which is requires less computation. This naturally leads to the question of the accuracy of this new model. As can be seen in Fig. \ref{fig:figure2} the new definition of entropy still leads to accurate identification of network structure. This new definition also fits into the general framework of entropy which was developed above, allowing us to apply the oMII algorithm to the data. 

\begin{figure}[htbp] 
\includegraphics[width=0.49\textwidth]{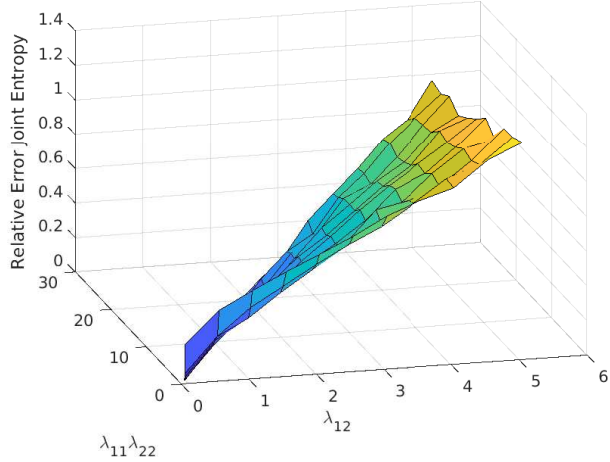} 
\caption{The relative error in the joint entropy calculation between the joint entropy calculated through truncation and the joint entropy calculated by our approximation. It is clear that when both $\lambda_{12}$ and $\lambda_{11} \lambda_{22}$ are small, the relative error is small. Thus we expect this approximation to work well when all of the estimated rates are small. In practice we find that when scaling the rates to be in $[0,1]$ we get good results, regardless of how high the true rates were. \label{fig:figure1}}
\end{figure} 
\subsection{Network Structure and Inference}

In a gene interaction network, understanding how future treatments could be developed, especially in the cases where more than a single gene may be implicated in a disease, may help in designing targeted for therapies. Genes interact with outcome such as disease reduces to a network inference problem.  We do not assume apriori knowledge of the underlying network structure, but instead we have data describing time series of evolving stochastic processes at each of the states, related to each individual  gene.
The network is stated as a graph $\mathcal{G}$ defined as a set of vertices $\mathcal{V} \subset \mathbb{N}$ and edges $\mathcal{E} \subset \mathcal{V} \times \mathcal{V}$, $\mathcal{G} = \{\mathcal{V}, \mathcal{E} \}$. 
Note that $|\mathcal{V}| =  n$ denotes that there are $n$ vertices (or nodes) in $\mathcal{G}$, by the cardinality, $| \cdot |$,  of a set.
The adjacency matrix $\mathcal{A} \in \mathbb{N}_0^{n \times n}$ is a convenient way to encode a graph,  
\begin{equation}
\begin{cases}
\mathcal{A}_{ij} = 1 \ \mbox{if} \ (i,j) \in \mathcal{E},  \\
\mathcal{A}_{ij} = 0 \ \mbox{otherwise}. \label{eq:adjacency}
\end{cases}
\end{equation}
When a system has a graph structure it is often referred to as a network. The adjacency matrix then encodes the network structure of the system. Our goal is to estimate network structure $\hat{\mathcal{A}}$  closely as  possible to the true network structure $\cal{A}$, that is we want, $\sum_{i,j} |\cal{A}-\hat{\mathcal{A}}|,$ to be as small as possible (ideally $0$). We would also like for this to be accomplished with as little data $(t)$ as possible, since we are often limited in the amount of real world data we receive. Our estimation of the network structure relies on nodes sharing information with one another. Thus $\hat{\mathcal{A}}$ may be thought of as which nodes are directly communicating with one another, rather than strictly being the physical structure. In our previous work, \cite{sun2014,sun2015}, we proved that under mild hypothesis, the multi-variate stochastic evolving by coupling on a complex network can be derived perfectly by optimal causation entropy (oCSE), errors arising from estimation issues such as model entropies of observations from various distributions, and finite data effects,  but the information network structure align accurately in most situations.

In the first example demonstration of our methods, we benchmark with synthetic simulated   by the multivariate Poisson model, Eqs. \ref{eq:multiPoiss_a} - \ref{eq:multiPoiss_c}. To explicitly incorporate the adjacency matrix $\mathcal{A}$ and noise $E$ as shown in \cite{allen2013,gallopin2013}, consider as:
\begin{align}
X = BY + E, \\
B =[ I_n;  P \odot (1_n tri(\mathcal{A})^T)]
\end{align}
where $I_n$ is the $(n \times n)$ identity matrix, $P \in \mathbb{N}^{n \times (m-n)}_0$ is a permutation matrix with exactly $n$ ones per row $\odot$ represents the Hadamard product (componentwise multiplication of same sized arrays). $1_n \in \mathbb{N}^{n \times 1}$ is the vector of all ones, and $tri(\mathcal{A}) \in \mathbb{N}^{\frac{n(n-1)}{2}\times 1}_0$ denotes the vectorized upper triangular portion of the adjacency matrix, and  $E \in \mathbb{N}_0^{n \times t}$. We have established in previous discussion that there is no analytical solution for the entropy of the multivariate Poisson, instead an approximation has been made. Since the Poisson distribution resembles the Gaussian distribution often the latter is assumed for estimates, we thus compare the performance of oMII assuming both distribution types. Fig. \ref{fig:figure2} shows that the oMII method, but even using the rough Gaussian best estimates of entropies, nonetheless does reasonably well finding the true edges with a high true positive rate (TPR).  This is contrasted to network inference based on other entropy estimators, including the nonparametric kNN method, GLASSO, both of which are discussed below, and also the Poisson estimator developed here. However, the Gaussian oMII finds the edges at the expense of a much larger false positive rate (FPR). Specifically, define TPR and FPR as follows: let $\mathcal{G} = \{\mathcal{V},\mathcal{E} \}$ be the true network structure and $\hat{\mathcal{G}} = \{\hat{\mathcal{V}},\hat{\mathcal{E}} \}$ be the estimated network structure.  Then:
\begin{equation}
\mbox{TPR} = \frac{|\mathcal{E} \cap \hat{\mathcal{E}}|}{|\mathcal{E}|}, \label{eq:truepositive}
\end{equation}
and
\begin{equation}
\mbox{FPR} = \frac{|\hat{\mathcal{E}} \ \textbackslash \ \mathcal{E}|}{|\mathcal{E}|}. \label{eq:falsepositive}
\end{equation}
In this case $\textbackslash$ represents set subtraction. Note that from this definition $0 \leq \mbox{TPR} \leq 1$ while $\mbox{FPR} \geq 0$.
\section{Results}
\begin{figure*}[htbp] 
\includegraphics[width=0.95\textwidth]{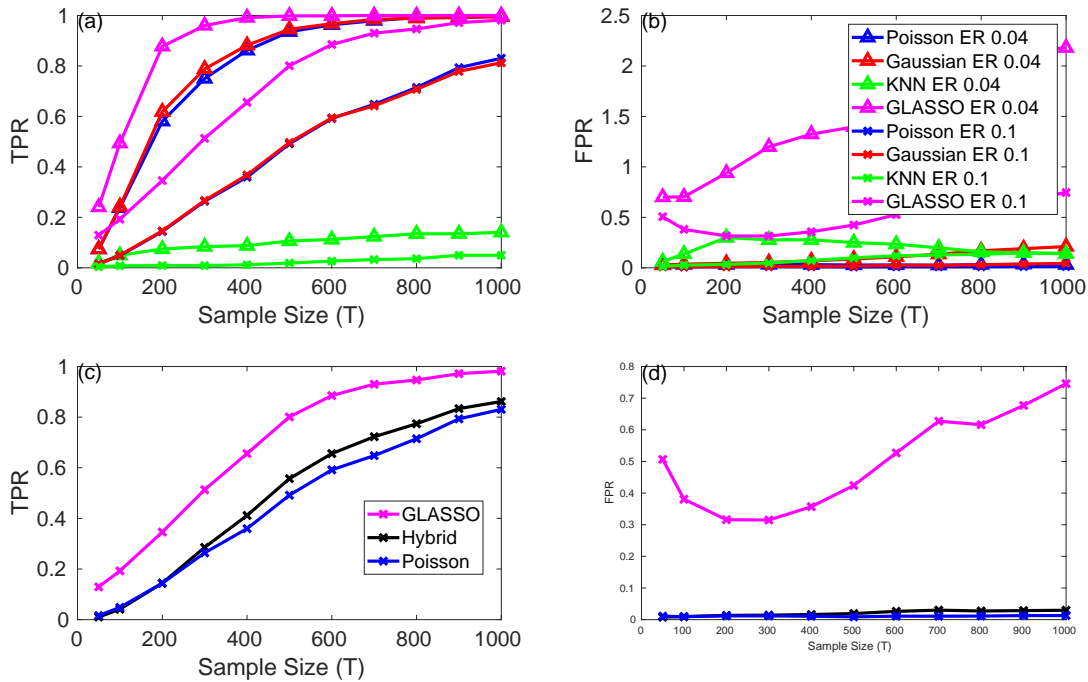}
\caption{True Positive and False Positive Rates for several test methods on ER graphs of two different levels of sparsity. Erd\H{o}s-R\'enyi (ER) graphs with triangles  for a 50 nodes graph with strong sparsity due to $p = 0.04$, and the x's for 50 nodes ER graphs with due to  denser  $p = 0.1$. The magenta lines represent GLASSO, the blue lines represent the Poisson oMII, the red lines represent the Gaussian oMII, and the green lines represent the KNN oMII. In (a) the true positive rate (TPR) is shown for different sample sizes, each point is averaged over 50 realizations of the network dynamics. In (b) the false positive rate (FPR) is shown. Clearly GLASSO finds more  true edges, but at the expense of a significantly higher false positives. In fact, for the highly sparse ER network GLASSO finds 3 times as many edges as actually exist in the network with 1000 data points. The FPR increases with data set. As can be seen the Gaussian oMII performs as well as Poisson oMII in TPR with the KNN performing poorly,  but the Poisson oMII significantly outperforms all other methods in terms of FPR. It appears that the Poisson oMII is the only method that converges to the true network structure with increasing sample size. (c) Comparing TPR between GLASSO, the hybrid method and Poisson oCSE. The hybrid method has an increased TPR relative to Poisson oCSE. (d) The FPR increases slightly for the hybrid method, but is substantially lower than GLASSO.\label{fig:figure2}}
\end{figure*}

We compare the performance of several methods on simulated data sets, including various types of oMII, as well as GLASSO \cite{friedman2008}. Unlike oMII, which involves conditional mutual information as its engine, GLASSO involves maximizing the log-likelihood provided in Eq. \ref{eq:GLASSO} over values of a regularization parameter $\rho$,
\begin{equation}
\mathcal{L}(X,\rho) =   \mbox{log}(\mbox{det}(\hat{\mathcal{A}}))-\mbox{trace}(\mbox{Cov}(X)\hat{\mathcal{A}})-\rho ||\hat{\mathcal{A}}||_1. \label{eq:GLASSO}
\end{equation}
A common method for the choice of $\rho$ is maximazation of the Bayesian information criterion (BIC). We utilize $1000$ log-spaced values of $\rho$ in $[10^{-2}, 1]$ which varies $\hat{\mathcal{A}}$ between a complete network to a completely disconnected network with zero edges. Following \cite{gallopin2013}, first we use a box-cox transformation of the Poisson distributed data, to make the data more Gaussian like, prior to using GLASSO. The box-cox transformation of a random variable $z$ is 
\begin{equation}
    bc(z|\gamma) = \begin{cases}
    \frac{z^\gamma -1}{\gamma} \ \text{if} \gamma \neq 0
    \\
    \mbox{log}(z) \ \text{if} \gamma = 0.
    \end{cases}
\end{equation}
GLASSO results are shown in Fig.~\ref{fig:figure2}.
\paragraph*{}
The Poisson oMII method is tested on data simulated as described in the section above. In Fig. \ref{fig:figure2} each data point is averaged over 50 realizations of the network dynamics. Two different  Erd\H{o}s-R\'enyi (ER) graph types are used, one with $p = 0.04$ and one with $p = 0.1$. The parameter $p$ in an ER graph controls the sparsity of the graph, thus the graphs with $p = 0.1$ will have considerably more edges on average than graphs with $p = 0.04$. For these simulations $n = 50$ was chosen. The rates were chosen to be $\lambda_{ij} = 1 \ (\forall \ i,j)$ and $E_i \sim \mbox{Poisson}(0.5) \ (\forall \ i)$ where $E_i \in \mathbb{N}^{t \times 1}_0$ are the columns of $E$. This is the high SNR scenario from \cite{allen2013}. To estimate the rates, we simply use correlation between all pairs in the data. We note that this differs from above where we utilized the covariance matrix. Using correlation rather than covariance guarantees the calculated rates will be relatively small since the values of correlation do not exceed $1$ in absolute value, this allows the estimated rates to stay in the small relative error regime shown in Fig. \ref{fig:figure1}. The correlation matrix then gives us all of the off diagonal rates $\lambda_{ij} \ (i \neq j)$ and to obtain the rates $\lambda_{ij} \ (i = j)$ we can see from Eq. \ref{eq:CovarianceMat} that we simply need to subtract the sum of the non-diagonal elements from the diagonal elements. That is if we let
\begin{equation}
\mbox{Corr}(X) =
\begin{bmatrix}
e_{11}  & \lambda_{12} & \cdots & \lambda_{1n} \\
\lambda_{12} & e_{22} & \cdots & \lambda_{2n} \\
\vdots & \vdots & \cdots & \vdots \\
\lambda_{n1} & \lambda_{n2} & \cdots  & e_{nn}  
\end{bmatrix}, \label{eq:CorrMat}
\end{equation}
then $\lambda_{ii} = e_{ii} - \sum \limits_{j\neq i}\lambda_{ij}$. In Fig. \ref{fig:figure2} it can be seen that in terms of TPR all of the methods perform quite well with the exception of the KNN version of oMII which exhibits poor performance across all examined sample sizes, likely due to slow convergence. In fact, for networks with few connections the poorest performing method in terms of TPR is the Poisson oMII method, with the best performing method being GLASSO. However GLASSO produces a very high FPR, in fact GLASSO finds {\it more} false positives than there are total edges in the true network, thus producing an FPR of greater than $1$. By contrast both Gaussian and Poisson versions of oMII produce significantly lower FPR and the Poisson oMII produces the lowest rate of FPR across all sample sizes. It should be noted as well that the FPR of the Poisson version of oMII maintains an approximately constant level across all sample sizes, while the Gaussian version of oMII has an increasing FPR with sample size. For the denser networks, which had an expected average degree of $5$, as expected all methods had a decreased TPR for low sample size. The FPR also fell for all methods due to the larger denominator (more edges). The conclusions remain the same for both network densities.

\begin{figure*}[htbp] 
\includegraphics[width=1.1\textwidth]{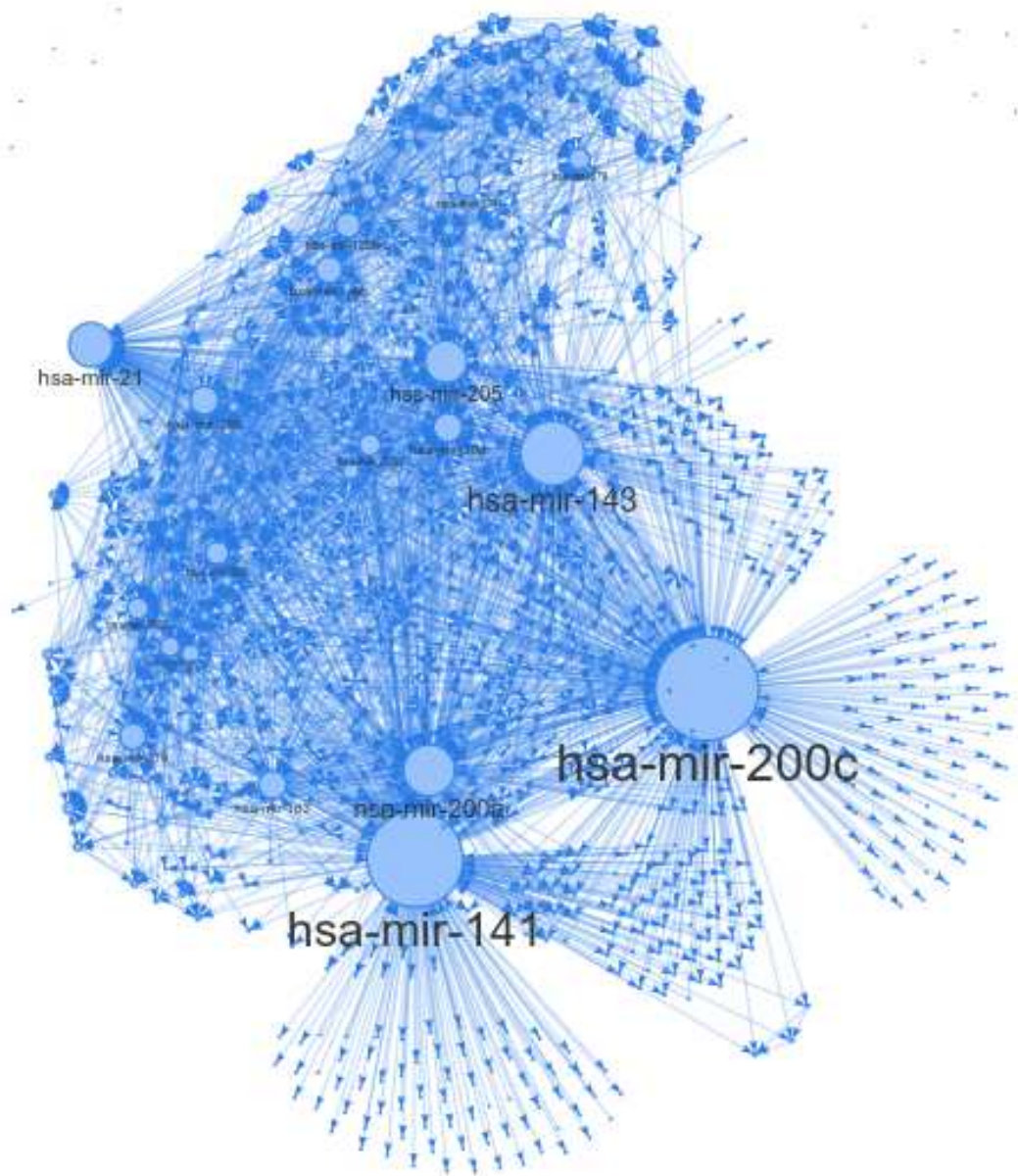}
\caption{Example network generated by the hybrid oMII algorithm. Nodes and text are sized relative to the out degree of the node. The nodes with largest out degree have previously been connected with breast cancer. \label{fig:figure3}}
\end{figure*}

\paragraph*{}
We now examine data derived from breast cancer patients who have been screened for different  micro RNA's (miRNA’s)  occurrence counts of  is analyzed by the Poisson oMII method featured in this paper. These data sets are publicly available at  https://portal.gdc.cancer.gov website, described as  TCGA-BRCA sequencing miRNA.
  In this case, $t = 1207$ and $n = 1881$ different miRNA samples are available. Of these, $1881$ miRNA's $\approx 1000$ pass the two sample Kolmogorov-Smirnov (KS) \cite{lilliefors1967} test comparing to the Poisson  distribution, to confidence level $\alpha = 0.05$. The remaining $\approx 900$ miRNA data were then scaled as follows: 
\begin{equation}
x_i^*  \in \mathbb{N}^{1207 \times 1}_0= \lfloor \frac{x_i}{<x_i>} \rfloor .
\end{equation}
The notation, $< \cdot >$ represents the mean and $\lfloor \cdot \rfloor$ componentwise, to integers. The scaled data is well fitting, again by KS-test, to a negative binomial distribution, with only $\approx 200$ failing as both Poisson and negative binomial. Recall that the Poisson distribution is a special case of the negative binomial distribution, since:
\begin{equation}
P_{NegBin}(k) = {k+r-1 \choose k} \lambda^k (1-\lambda)^r. \label{eq:NegativeBinomial}
\end{equation}
In the limit, $r \rightarrow \infty$ in Eq. \ref{eq:NegativeBinomial} it is easy to see that the term $(1-\lambda)^r \rightarrow e^{-\lambda}$, and rewriting ${k+r-1 \choose k} = \frac{(k+r-1)!}{k!(r-1)!} \rightarrow \frac{1}{k!}$. Combining these facts,  as $r \rightarrow \infty$, the negative binomial distribution limits to a  Poisson distribution.
\paragraph*{}
Given that the majority of this miRNA data is distributed as scaled negative binomial (the Poisson data also can be fit as negative binomial) we must interpret the results of with caution especially in light of the results shown in Fig.~\ref{fig:figure2}. The results of the application of the Poisson oMII still are interesting, especially in light of the fact that the negative binomial distribution can be viewed as a compound Poisson distribution \cite{anscombe1950,bissell1972}. To obtain the networks shown in Fig. \ref{fig:figure3} we first restricted the data to having a minimum of $>100$ total counts,this was to avoid including data that had zero variation or near zero variation. This restriction left us with $1072$ miRNA's, oMII was then used to analyze the remaining miRNA data without any further pre-processing, which resulted in the network shown in Fig. \ref{fig:figure3}. The network has many miRNA's which are non-interacting, however there is a large weakly connected component. Focusing on the nodes which are members of the largest weakly connected component (LWCC) we found that many miRNA's that have been previously identified as up or down regulated in breast cancer end up in this component, this component included most of the miRNA's listed in Table 1 of \cite{iorio2005}. The miRNA's which land in the LWCC will be labeled as {\it interesting} miRNA's for brevity. 
\paragraph*{}
Focusing on this set of 656 miRNA's, the plot of Fig. \ref{fig:figure3} focuses in on this component by sizing the nodes relative to their out degree. The nodes with no out degree are so small that they are difficult to see in the figure, while the nodes with largest out degree are prominent. A feature of this network is that there are miRNA's that are "drivers" of the network, in that they have much larger out degree than the majority of other nodes.  We list the top 20 miRNA's in order of their centrality based on out degree, betweenness centrality and eigenvector centrality in Table \ref{tab:BigTable}. For all three measures the top 4 miRNA's are identically ordered, all 4 of which have been noted for a prominent role in breast cancer \cite{iorio2005,lim2013,antolin2015,thammaiah2016,medimegh2014,tanic2015} and they seem to be the main drivers. This suggests that it may be possible to target a small number of miRNA's for some desired behavior of the system of miRNA's in drug development.
\begin{table}
\begin{tabular}{ |c|c|c| } 
 \hline
 Out Degree & Betweenness Centrality & Eigenvector Centrality \\ 
 Mir-200c & Mir-200c & Mir-200c  \\ 
 Mir-141 & Mir-141 & Mir-141 \\ 
 Mir-143 & Mir-143 & Mir-143 \\
 Mir-200a & Mir-200a & Mir-200a \\
 Mir-21 & Mir-205 & Mir-205 \\
 Mir-205 & Mir-21 & Mir-21 \\
 Mir-30a & Mir-26b & Mir-30a \\
 Mir-26b & Mir-30a & Mir-183 \\
 Mir-183 & Mir-183 & Mir-26b \\
 Mir-199b & Mir-199b & Mir-326 \\
 Mir-210 & Mir-125b-2 & Mir-200b \\
 Mir-125b-2 & Mir-134 & Mir-210 \\
 Mir-134 & Mir-326 & Mir-125b-2 \\
 Mir-326 & Mir-3607 & Mir-199b \\
 Mir-200b & Mir-379 & Mir-429 \\
 Mir-379 & Mir-210 & Mir-32 \\
 Mir-3607 & Mir-1976 & Mir-3607 \\
 Mir 429 & Mir-150 & Mir-134 \\
 Mir-32 & Mir-203a & Mir-766 \\
 Mir-337 & Mir-100 & Mir-100 \\
 \hline
\label{tab:BigTable}

\end{tabular}
\caption{The top 20 genes discovered from the hybrid method in terms of: out degree, betweenness centrality, and eigenvector centrality. All of these genes have been linked to breast cancer by previous studies.}
\end{table}
\section{Conclusion}
In this paper we have given an approximation to the mutual information of a multivariate Poisson system. We have shown through numerical experiments that this approximation works efficiently, and the results of network estimation indicate that the approximation is justified. We have also developed the oMII (and by extension the oCSE) algorithm for computation of the causation entropy of a Poisson system based on the joint entropy approximation discussed above. We have shown that this model is superior to simply assuming the data is Gaussian, which is likely related to the strange behavior of the marginals in a Poisson sytem, as we have outlined above. The Poisson oMII algorithm also significantly outperforms the nonparametric KNN version of oMII. Finally, we have applied the Poisson oMII algorithm to a breast cancer miRNA expression count dataset, which has produced potentially interesting insights into the network of miRNA's as it relates to breast cancer. Our network inference on the breast cancer miRNA network has shown that there is a relationship between the highest variance (in expression values) of miRNA's. There seems to be unidirectional connections between these miRNA's, with certain miRNA's taking on the role of drivers in the network. This may suggest a future course of action for future drug development.

\section{Acknowledgements}
E.B. was supported by the Army Research Office (N68164-EG) and, J.F. and E.B. were supported by DARPA.

\section{Appendix}
Below we offer proof of Eq. \ref{eq:CovarianceMat}.
\begin{proof}{Covariance of the multivariate Poisson}

In the model presented in Eqs. \ref{eq:multiPoiss_a}, \ref{eq:multiPoiss_b}, \ref{eq:multiPoiss_c}, we can see that:
\begin{align}
x_1 = y_{11} + y_{12} + ... + y_{1n} \label{eq:covModel_a} \\ \nonumber
x_2 = y_{12} + y_{22} + ... + y_{2n} \\ \nonumber
\vdots \\ \nonumber
x_n = y_{1n} + y_{1n}  + ... + y_{nn}
\end{align}
Without loss of generality we will look at the pair $(i =1, j = 2)$. In this case we see that the covariance between this pair of random variables is defined:
\begin{equation}
\mbox{cov}(x_1,x_2) = \mathbb{E}[x_1 x_2] - \mathbb{E}[x_1]\mathbb{E}[x_2], \label{eq:covModel_b}
\end{equation}
Considering Eqs. \ref{eq:covModel_a}, \ref{eq:covModel_b} and noting $y_{12} = y_{21}$, we have:
\begin{align}
&\mbox{cov}(x_1,x_2) =\mathbb{E} \left[y_{12}^2 + \sum \limits_{\substack{i = 1 \\ i \neq 2}}^n  \sum \limits_{j = 2}^n y_{1i} y_{2j}\right] \label{eq:covModel_c} \\ \nonumber 
&- \mathbb{E}[y_{11} + y_{12} + ... + y_{1n}] \mathbb{E}[y_{12} + y_{22} + ... + y_{2n}]
\end{align}
Because the expectation is a linear operator,Eq. \ref{eq:covModel_c} can be expressed as:
\begin{align}
& \mbox{cov}(x_1,x_2) = E[y_{12}^2] +  \mathbb{E}\left[ \sum \limits_{\substack{i = 1 \\ i \neq 2}}^n \sum \limits_{j = 2}^n y_{1i} y_{2j} \right] - \\
\nonumber & \left(\mathbb{E}[y_{12} ]+ \sum \limits_{\substack{i = 1 \\ i \neq 2}}^n \mathbb{E}\left[ y_{1i} \right]\right)  \left(\mathbb{E}[y_{12} ]+ \sum \limits_{j = 2}^n \mathbb{E}\left[ y_{2j} \right]\right).
\end{align}
From the independence of each $y_{ij}$ the covariance can thus be expressed:
\begin{align}
 \mbox{cov}(x_1,x_2) = & \mathbb{E}[y_{12}^2] + \sum \limits_{\substack{i = 1 \\ i \neq 2}}^n \sum \limits_{j = 2}^n \mathbb{E}\left[  y_{1i} y_{2j} \right] - \\
\nonumber & \mathbb{E}^2[y_{12} ] -\sum \limits_{\substack{i = 1 \\ i \neq 2}}^n \sum \limits_{j = 2}^n \mathbb{E}\left[  y_{1i} y_{2j} \right] = \\ \nonumber
&\mathbb{E}[y_{12}^2] - \mathbb{E}^2[y_{12} ] = \\ \nonumber
& \mbox{Var}(y_{12}) .
\end{align}
Since $y_{12}$ is independent Poisson and from the variance of an independent Poisson random variable $\mbox{Var}(y_{12}) = \lambda_{12}$. Applying this to each $i,j (i \neq j)$ pair gives the desired covariance structure.
\end{proof}

\end{document}